\documentclass[%
 reprint,
 amsmath,amssymb,
 aps,
]{revtex4-2}

\usepackage{graphicx}
\usepackage{dcolumn}
\usepackage{bm}
\usepackage[utf8]{inputenc}
\usepackage{CJKutf8}
\usepackage{float}

\begin{document}

\preprint{APS/123-QED}

\title{Modeling meteorite craters by impacting melted tin on sand}

\author{H. Y. Huang$^{1}$, P. R. Tsai$^{1}$, C. Y. Lu$^{1}$, H. Hau$^{1}$, Y. L. Chen$^{2}$, Z. T. Ling$^{3}$, Y. R. Wu$^{4}$ and Tzay-Ming Hong$^{1\ast}$}\

\affiliation{%
$^1$Department of Physics, National Tsing Hua University, 
Hsinchu, Taiwan 30013, Republic of China}
\affiliation{%
$^2$Department of Mechanical Engineering, National Taiwan University, 
Taipei, Taiwan 10617, Republic of China}
\affiliation{%
$^3$Department of New Media Art, Taipei National University of the Arts, Taipei, Taiwan 110024, Republic of China}%
\affiliation{%
$^4$National Hsinchu Senior High School, Hsinchu, Taiwan 30013, Republic of China}%




\date{\today}

\begin{abstract}
 To simulate the heated exterior of a meteorite, we impact a granular bed by melted tin. The morphology of tin remnant and crater is found to be sensitive to the temperature and solidification of tin. By employing deep learning and convolutional neural network, we can quantify and map the complex impact patterns onto network systems based on feature maps and Grad-CAM results. This gives us unprecedented details on how the projectile deforms and interacts with the granules, which information can be used to trace the development of different remnant shapes. Furthermore, full dynamics of granular system is revealed by the use of Particle Image Velocimetry.
 Kinetic energy, temperature and diameter of the projectile are used to build phase diagrams for the morphology of both crater and tin remnant. In addition to successfully reproducing key features of simple and complex craters, we are able to detect a possible artifact when compiling crater data from field studies. 
 The depth of  craters from high-energy impacts  in our work is found to be independent of their width. However, when mixing data from different energy, temperature and diameter of  projectile, a bogus power-law relationship appears between them. Like other controlled laboratory researches, our conclusions have the potential to  benefit the study of paint in industry and asteroid sampling missions on the surface of celestial bodies.

\end{abstract}

\maketitle


\section{Introduction}

Meteorite impact events have appeared in many disaster movies. However, Chelyabinsk meteor impact event and 2019 OK asteroid sweeping past the Earth remind us that meteorite impacts remain a menace to life on earth. Since the days of Galileo, geologists and astronomers have accumulated a lot of knowledge on the meteorite craters via telescope and field studies. 
This information can help us not only prepare for and mitigate potential risks posed by future impacts, but also understand the history and geology of habitable exoplanets.  Besides, impact craters provide valuable insights into the formation and location of recoverable mineral resources both on Earth and beyond.

Starting roughly 30 years ago, scientists began to smash various projectiles on a granular bed to simulate the impact event \cite{cite1,cite2,cite3}. This endows precious information, esp. on how the impact energy $E$, proportional to drop height $h$, affects the width $w$ and depth $d$ of crater. Roughly proportional to $d$, $w$  scales as $E^\alpha$ with $\alpha =0.25$ \cite{cite4} or 0.17 \cite{cite5} depending on whether a hard (e.g., steel ball) or  deformable (water droplet)  bullet is used. Meanwhile, it was found that some specific features, e.g., the  central peak in complex craters, are better reproduced by adopting a granular projectile \cite{cite6}. 
As opposed to field studies, knowledge accumulated from controlled laboratory research  like these also benefits the study of soil erosion in agriculture \cite{cite7,cite8}, paint in industry \cite{cite9}, and asteroid sampling missions on the surface of celestial bodies \cite{cite10,cite11}.  It is thus important to continue lab experiments of impacting granular beds.

\begin{figure*}[ht]
\centering
\includegraphics[scale=0.35]{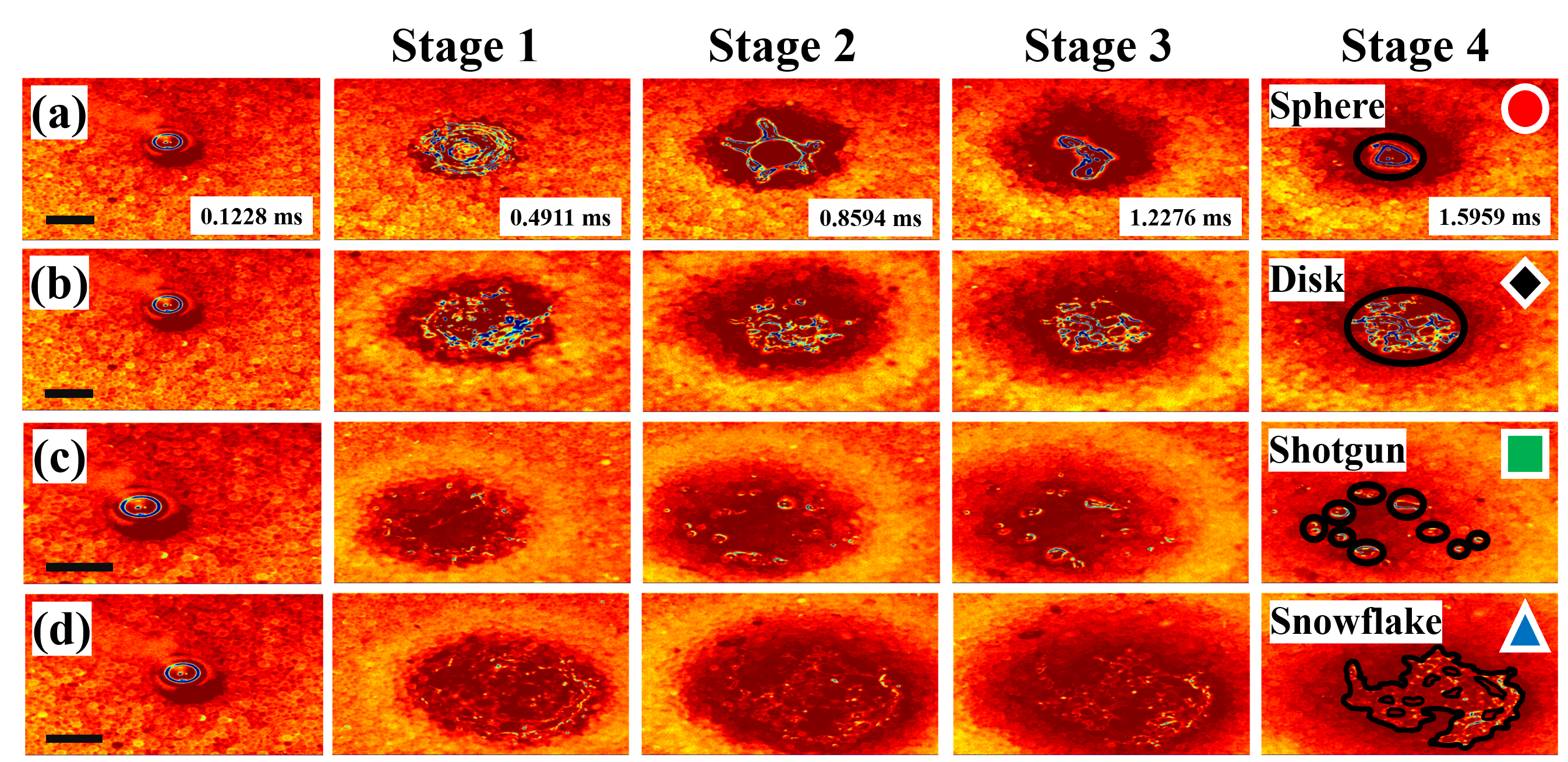}
\caption{ (a$\sim$d) Snapshots of melted tin of $R=6.4$ mm at  380 ℃ impacting  sand  for $h$ = 100, 250, 400, and 700 mm. Scale bars: 6.0 mm. 
False colors are used as an aid in visualizing features.
The shape of tin remnant, highlighted by dark solid lines, is  categorized into sphere, disk, shotgun, and snowflake, as labelled by red circle, black rhombus, green square, and blue triangle.}
\label{fig1}
\end{figure*}

 The important role of heat transfer and solidification for melted tin impacting on a solid surface has been discussed  \cite{naturephys} to gain insight on thermal spraying and additive manufacturing, and extreme ultraviolet lithography in chip production. In this Letter we adopt melted tin droplets as the bullet to better simulate the high-temperature environment typical of meteorite impact events.
Not limited to  $h$ in Fig. \ref{fig1}, the tin remnant also changes its shape with temperature $T$. This information will be summarised in diagrams and shown to be correlated to the morphology of crater and tin remnant. As in previous studies \cite{cite12},  we  also investigate $w(E)$ and $d(E)$ to see whether the solidification of melted tin will create any new feature. Particle image velocimetry (PIV)  and image processing are employed to analyze the flow field of granules, while image processing enables us to project  the  tin shape  on a 2-D plane and witness its deformation in real time. Additionally, we  use deep learning - convolutional neural network (CNN) and reverse backpropagation Grad-CAM to detect features of tin during the impact process and discuss their network properties.

\begin{figure*}[ht]
\centering
\includegraphics[scale=0.33]{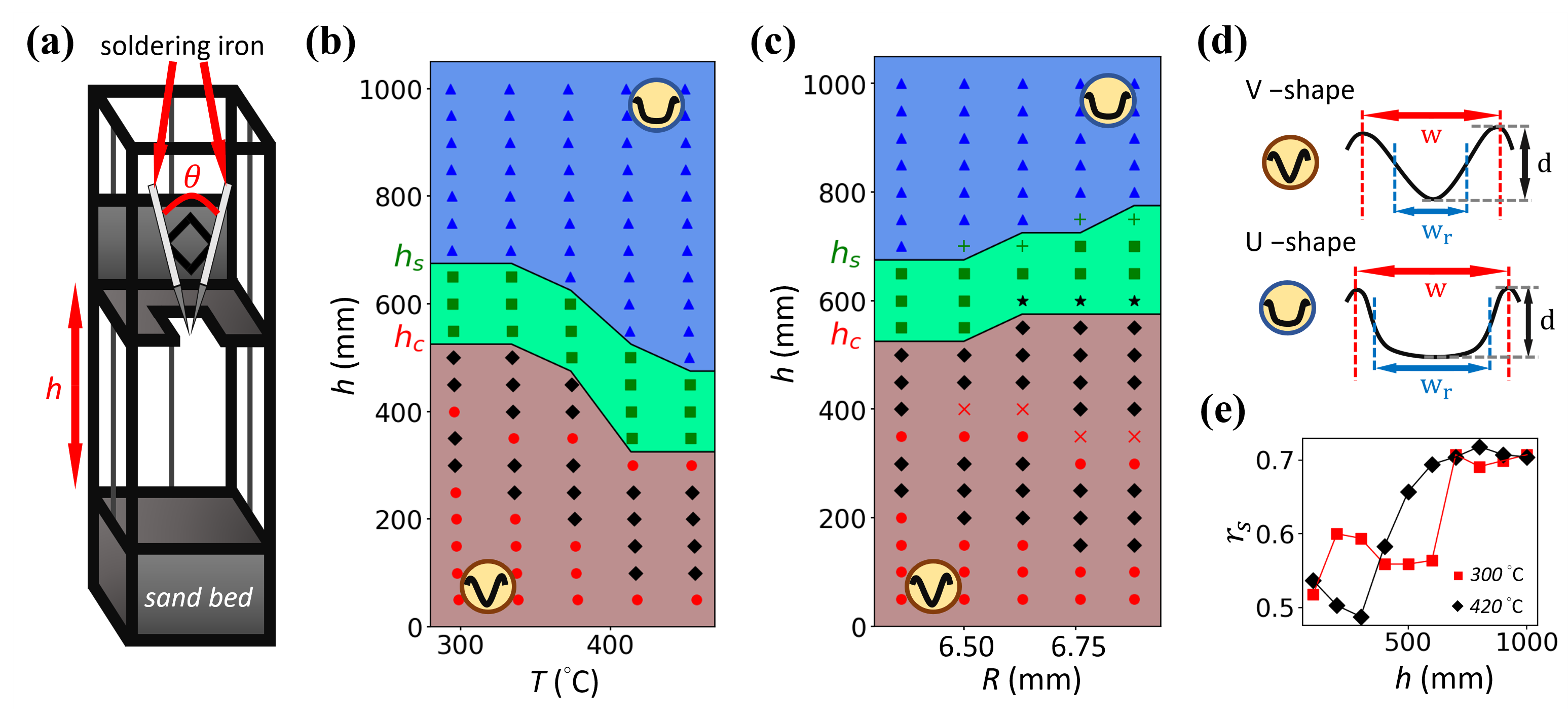}
\caption{(a) Schematic of experimental setup. (b) Phase diagram with $h$ and $T$ as axes shows the morphology of impact crater created by melted tin of $R$ = 6.4 mm can mimic either letter U (in blue) or V (brown). Green area denotes an intermediate phase, and labels for the remnant shape follow those in Fig. \ref{fig1}. (c) Similar phase diagram with $h$ and $R$ at 380 ℃. (d) Schematic profile of U- and V-shape craters. (e) Shape ratio $r_s$ vs $h$ at 300 and 420 °C, represented by red square and black rhombus.}
\label{fig2}
\end{figure*}

\section{Experimental Setup}

As shown in Fig. \ref{fig2}(a), our setup  is typical of impact experiments \cite{cite13}, except that it comprises of an elevated station with two soldering irons controlled at $T$ = 300, 340, 380, 420, and 460 °C, respectively. The spacing and relative orientation between irons can be adjusted to produce tin droplets of mass  0.30, 0.32, 0.34, 0.36, and 0.38 g or equivalently 6.4, 6.5, 6.6, 6.8, and 6.9 mm in diameter $R$. For the height $h$ of release from 0.05 to 1 m, the decrease in temperature during free fall is negligible. To eliminate the effect of granular compaction, we follow the  practice of Walsh by reloading the sand and scraping off extra sands by a ruler to flatten its surface before each trial \cite{cite1}. The impacts are characterized by the inverse Frode number ${\rm Fr}^{-1}=gD_b/2v^2$, the ratio of gravitational to dynamic pressures \cite{cite1}. Here $g$ denotes the gravitational acceleration, $D_b$  the projectile  diameter, and $v$ the impact velocity. In our system,  ${{10}^{-3}\le {\rm Fr}^{-1}}\le{10}^{-2}$ overlaps with  ${{10}^{-6}\le {\rm Fr}^{-1}}\le{10}^{-2}$ for meteorite craters, which justifies the analogy to astronomical impact events \cite{cite3}.


\section{Experimental Results}

\begin{figure*}[ht]
\centering
\includegraphics[scale=0.35]{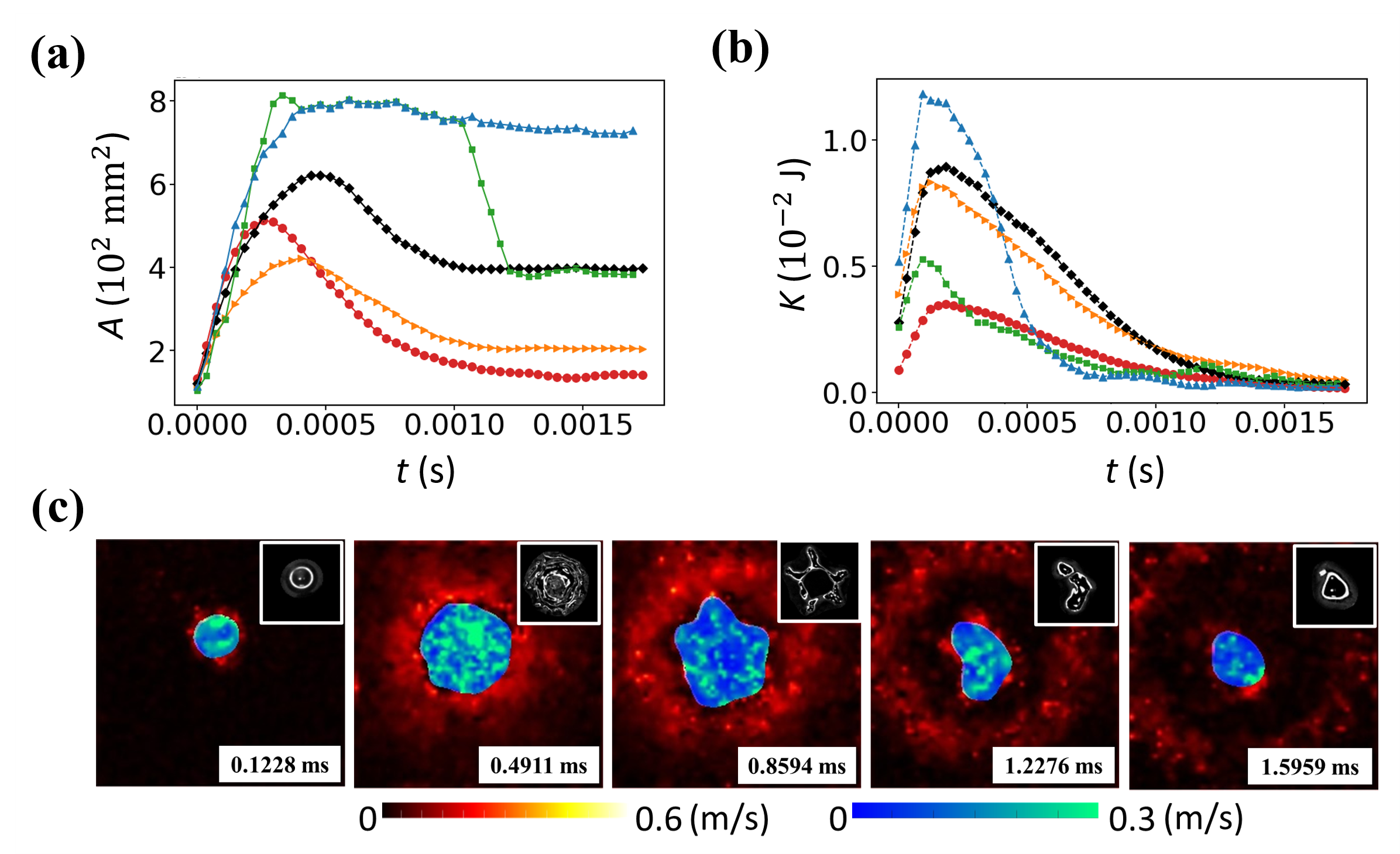}
\caption{(a) The top-view area $A$ of tin remnant versus time $t$ for sphere, sphere-II, disk, shotgun, and snowflake shapes is denoted by red circle, orange triangle, black rhombus, green square, and blue triangle, respectively. (b) The total kinetic energy $K$ of sand for different tin shapes is plotted as a function of $t$. (c) PIV results for snapshots show the evolution of melted tin with $R=6.4$ mm, 380 ℃ and  $h$ = 50 mm where velocity colorbars for sand on the left and tin on the right. Image of tin remnant after texture segmentation \cite{si} is inserted on the upper right corner of each photo.}
\label{fig3}
\end{figure*} 

The morphology of melted tin   in Fig.  \ref{fig1} is recorded by a high-speed camera of 8146 fps.
Irrespective of $h$, the process can be roughly separated into four stages based on the configuration of tin droplet: (1) circumference increases with time, but remains approximately circular, (2) irregular shape emerges, (3) its boundary retracts, and (4) static configuration is reached. As the impact energy increases, the final shape of tin remnant  can roughly be categorized into four shapes: sphere, disk, shotgun, and snowflake, as illustrated in Fig. \ref{fig1}.

 The sequence of appearance for these shapes can be understood physically. Since the distortion of tin melt raises its surface energy, we expect the aspect ratio of disk to increase with impact energy. As the thickness of disk reaches the threshold of Rayleigh-Plateau (RP) instability \cite{cite14}, clumping ensues and produces   many small and smooth tin balls in the stage-4 photo of Fig. \ref{fig1}(c), which mimics the shotgun pattern. As more kinetic energy is fueled, fast spreading of melted tin hastens the conduction rate of its heat to the sand bed. As a result, the transition from liquid to solid phase precedes and prohibits the  RP instability. This creates a splash-like \cite{cite15} irregular configuration in one piece, which we term snowflake in Fig. \ref{fig1}(d).  

\subsection{Phase Diagrams for Morphology}

\subsubsection{Tin Remnants}

Figure \ref{fig2}(b, c) show the phase diagrams for the morphology of both the crater and tin remnant as a function of parameters ($T$, $h$) and ($R$, $h$). 
At low impact energies or for $h$ = 50$\sim$250 mm, a higher $T$ renders the melted tin more susceptible to flattening upon impact.
At slightly higher energies  or $h$ = 400$\sim$500 mm, $h_c$ roughly marks the emergence of shotgun pattern at high temperatures, again because the fluid behavior is less restricted which allows the RP instability to enter the picture. Within the same scenario that high temperature promotes fluidity, the transition from shotgun to snowflake around  $h_s$ is analogous to the effect of increasing $h$ at high energies. We do not understand why two shapes can be observed in some intervals of $R$ in Fig. \ref{fig2}(c), as marked by crosses, daggers, and stars. 

\begin{figure*}[ht]
\centering
\includegraphics[scale=0.34]{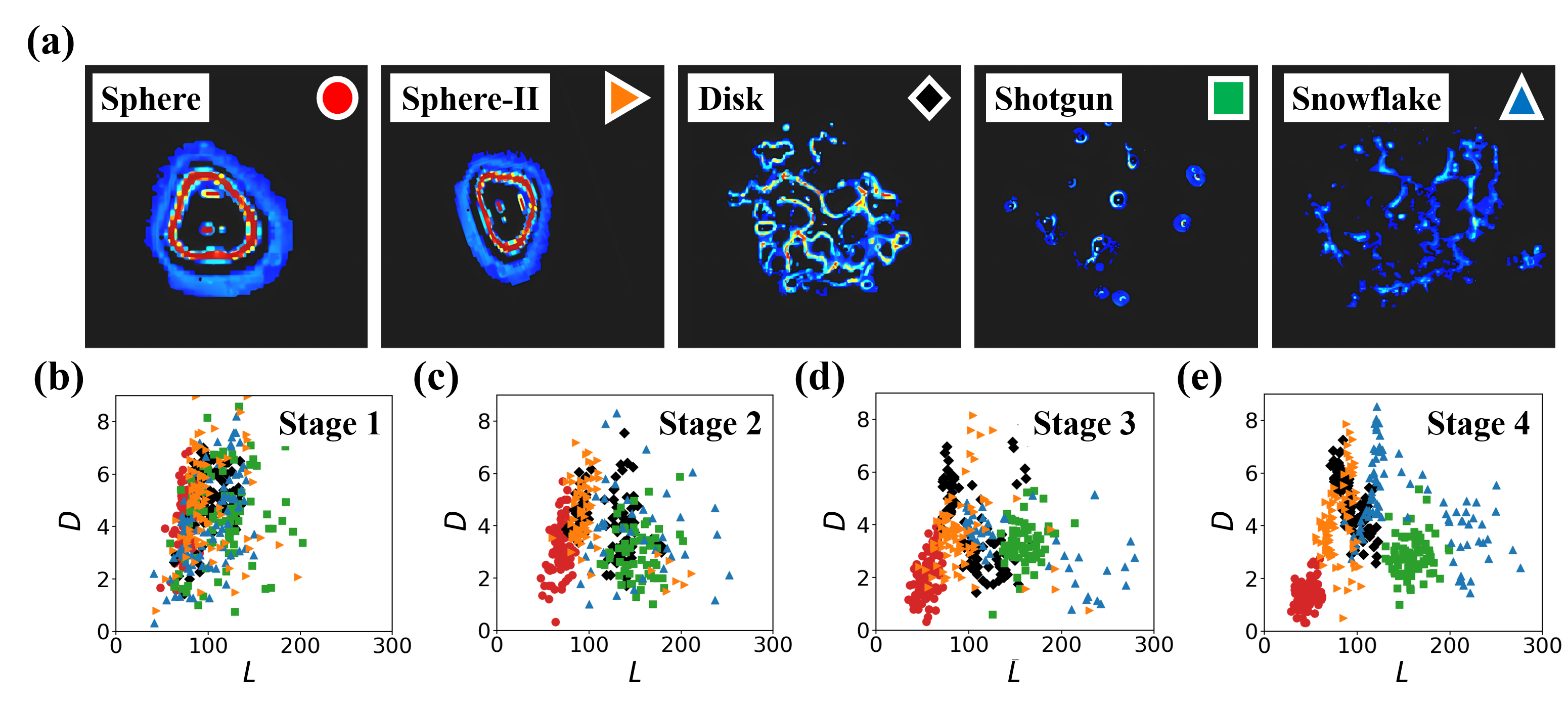}
\caption{(a) Grad-CAM results for different shapes of tin remnant.  Network density $D$ vs $L$ are shown in (b$\sim$e) for stage 1$\sim$4  respectively where the symbols are defined in (a). }
\label{fig4}
\end{figure*}

The trend of phase boundary in Fig. \ref{fig2}(c) for remnant shape at low energies mimics that in Fig. \ref{fig2}(b). It is because the dissipation rate per unit mass is proportional to the surface-area-to-volume ratio which for a sphere drops as $R$ increases. A larger melted tin thus gets to retain its malleability longer and better. This effect of preferring disk over sphere at low impact energy is similar to raising $T$. 
Ostensibly, the trend of $h_c$ and $h_s$ in Fig. \ref{fig2}(b) are both reversed in Fig. \ref{fig2}(c). Their physical explanations are in fact similar and consistent.  The reduction in thickness as we decrease $R$ favors the occurrence of RP instability at medium $h$, while making it easier to lose heat and solidify before RP occurs at high $h$ - both effects are similar to those by increasing $T$. 

For complex meteorite craters, due to extremely large mass,  more energy will transform into heat during the  impact process \cite{cite16}. As a result, the material on the surface of celestial bodies will change to liquid or gas phases. It was observed that the edge of these large craters is  so steep that it often collapses under its own weight and  concentrates at the bottom of craters \cite{cite17}. This is one of several mechanisms of the central peak in the majority of complex craters  \cite{cite18}. Note that the spherical shape reappears at higher-$h$ in Fig. \ref{fig2}(b, c), which is associated with a steeper impact crater in stage 2 and will be denoted by sphere-II. The appearance of this shape is always accompanied by a quasi-bump forms in the middle of crater as the sand on the wall and rim of the transient cavity collapses and collides at the center in stage 3. This mimics the central-peak structure in complex craters. 

\subsubsection{Crater}
We first try to reproduce the general observation that the complex craters tend to be more plate-like with $d/w = 0.05\sim 0.1$, as opposed to bowl-shaped with $d/w = 0.14\sim 0.2$ for simple ones \cite{cite18}. With laser displacement sensors and a homemade translation stage, we can obtain detailed information on the profile of our impact crater. In Fig. \ref{fig2}(b, c),  two critical heights, $h_c$ and $h_s$, are defined to manifest the shape change of tin remnant. Physically we expect the transition to leave its mark on the morphology of crater and, thus, introduce the shape ratio $r_s\equiv w_r/w$ to quantify the profile where $w_r$ is  the distance between inflection points as shown in Fig. \ref{fig2}(d). 
According to Fig. \ref{fig2}(e), it turns out that $h < h_c$ indeed corresponds to a smaller $r_s$, i.e., resembling V-shape or a bowl. On the other hand, U-shape or plate-like craters are more likely to occur at $h> h_s$. This is consistent with the distinctive shapes for simple and complex craters. 
We speculate that this dichotomy in shape has to do with the viscoelastic property of sand and soil \cite{cite2} - a high-energy collision triggers more elastic response that is more effective at halting the forward motion of tin and forcing it to spread out horizontally.

\begin{figure*}[ht]
\centering
\includegraphics[scale=0.33]{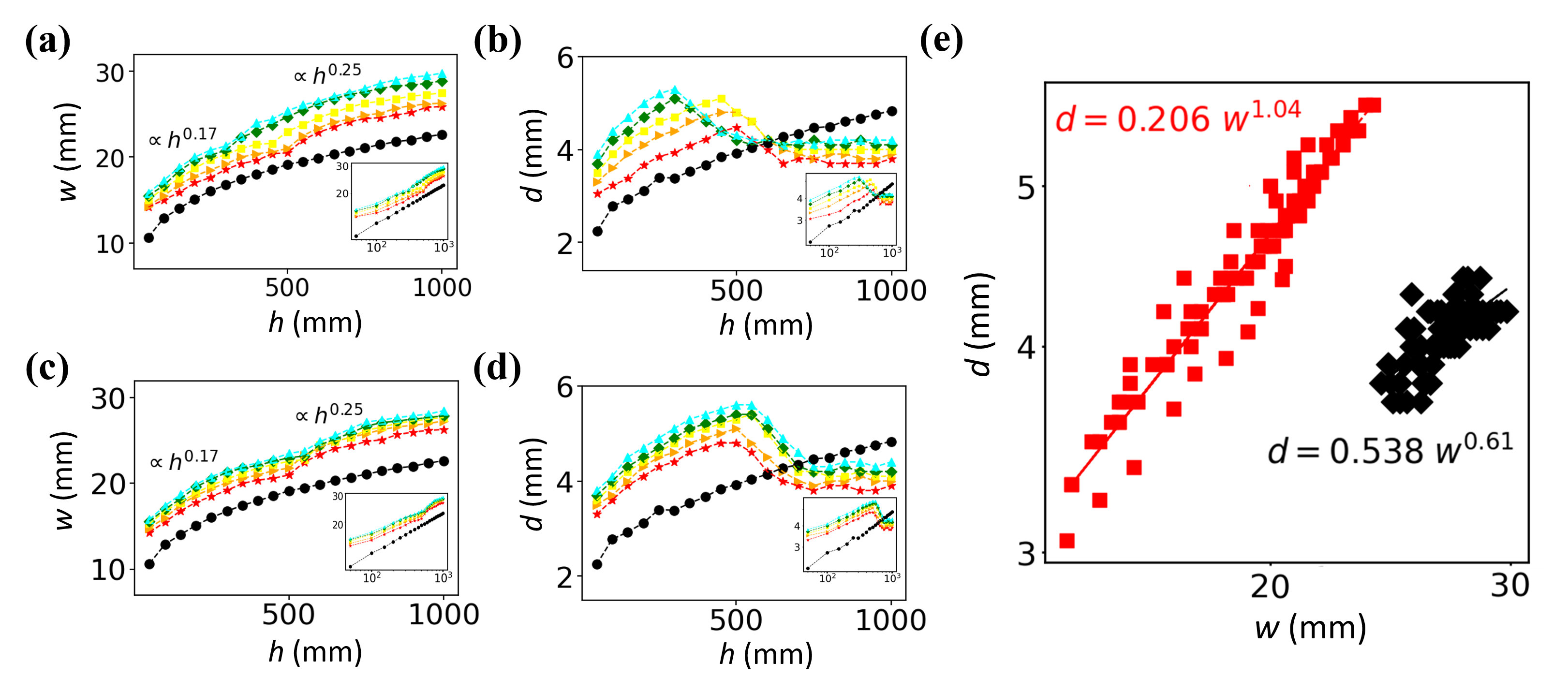}
\caption{(a) Crater width $w$ vs $h$ for steel balls (in black circle) and melted tin  with 300, 340, 380, 420 and 460 ℃ (red star, orange triangle, yellow square, green rhombus and cyan triangle). (b) Crater depth $d$ vs $h$ with the same symbols as (a). (c) $w$ vs $h$ for steel balls (black circle) and melted tin with 6.36, 6.5, 6.63, 6.76 and 6.88 mm (red star, orange triangle, yellow square, green rhombus and cyan triangle). (d) $d$ vs $h$ with the same symbols as (c). The insets in (a$\sim$d) show the same data in  full-log plots. (e) $d$ vs $w$ for V- (red square) and U-shape craters (black rhombus).}
\label{fig5}
\end{figure*}

\subsection{Deep-learning-augmented Analysis}
\subsubsection{Expanded Area Projection and PIV}

In order to confirm our speculations in Sec. III.A.1 regarding the different shapes of tin remnant, we employ image processing techniques and a high-speed camera to measure the top-view area $A$ of tin at successive times in Fig. \ref{fig3}(a). Note that the exhibition of a peak value for all lines signals the time beyond which the melted tin starts to retract. Two things worth mentioning. First, the maximum $A$ value of sphere-II is smaller than that of sphere. This is against our intuition because the former is dropped from a higher $h$. It turns out that sphere-II causes a steeper cavity in stage 2 which hinders the spread of tin. Second, there exists a plateau immediately following the peak for shotgun and snowflake. This is ascribed to the faster solidification due to their lesser thickness when spread out to a larger area. As for the sudden drop following the plateau for shotgun, it is due to the  breakup of melted tin into several small spheres.

Using PIV analysis and image processing, we can measure the speed of moving granules and calculate their total kinetic energy $K$, as shown in Fig. 3(b). Physically $K$ has to compete with the surface energy of melted tin for the energy converted from the gravitational potential during the impact.
The reason why shotgun allocates  lesser $K$ for sand than other shapes from a lower height is that the surface energy already gets the lion's share. Aside from the leftmost image that illustrates the initial impact, the successive images in Fig. 3(c) correspond to the four stages.

\subsubsection{Network of Tin Remnants}

Creating the equivalent network for the tin remnant allows us to  distinguish the highly complex pattern of tin remnants  quantitatively and more objectively. This is made possible by employing CNN and Grad-CAM \cite{cite20}.  The structure and learning curve of CNN can be found in Ref.\cite{si}. We adopt the features, used by Grad-CAM in distinguishing remnant shapes, to create distance-based networks \cite{si}. The Grad-CAM results are shown in Fig. 4(a).
 
Based on the network density $D$ and the average distance $L$ between nodes, we are able to not only pinpoint when the morphology starts to develop different shapes, but also quantify their relative proximity. According to Fig. 4(b$\sim$e), we learn that the shape branches off in stage 3 or 4. In contrast to Fig. 2(b,c) using $h$, $T$ and $R$ as parameters, the relative position  of shapes  in Fig. 4(e) is based on the network properties. The proximity of shapes, say, disk and sphere-II, disk and shotgun, and shotgun and snowflake, revealed by both figures is consistent. What is surprising is that
snowflake and sphere-II that are separated by a third shape in Fig. 2(b,c) turn out to share a very similar network, whiles the opposite happens to sphere and disk, i.e., bordering shapes in Fig. 2(b,c) turn out to drift apart in Fig. 4(e).



\subsection{Diameter and Depth of Crater}
Figure 5(a$\sim$d) reveals how the width $w$ and depth $d$ of craters are determined by the impact energy or equivalently the height $h$. 
Consistent with previous studies, $w(h)$ is found to obey a power-law relation, i.e., $w\propto h^\alpha$ \cite{cite21}. However, in contrast to $\alpha$=0.17 for water droplet \cite{cite22} and 0.25 for steel ball \cite{cite4}, both values occur in Fig. 5(a, c), depending on whether $h$ is smaller or greater than  $h_c$ as defined  in Fig. 2(b, c). This change of behavior was reminiscent of the observation for granular ball \cite{cite6}, except that the latter exhibits a discontinuous increase and the $\alpha$ on both sides are close to 0.25.

The $d\propto h^\beta$ is monotonically increasing with $\beta=1/3$ \cite{cite23} or 0.25 \cite{cite4} for steel ball and $\beta=0.17$ \cite{cite22} for water droplet, but exhibits a discontinuous drop for granular ball \cite{cite6}. In Fig. 5(b, d) for melted tin, $d(h)$ is not monotonic - increase at $h<h_c$, and decrease continuously when $h_c<h<h_s$ before finally saturate at $h>h_s$. Note that the data points in Fig. 5(b, d) for melted tin at  $h<h_c$  are higher than that for steel ball with the same weight. At first glance, this is counter-intuitive because melted tin ought to have less energy to dig the crater while expanding its area at the same time. The answer to this puzzle is that the crater is largely formed in stage 2 before melted tin has fully retracted to become a sphere or disk. It was checked \cite{si} that while the bottom of this V-shape tin sheet or, equivalently, $d$ for crater is deeper than  steer ball, their centers of mass are in fact at roughly the same depth in stage 2. This explanation is consistent with the fact that the discrepancy of $d$ for melted tin and steer ball increases with temperature.

By compiling the data in Fig. 5(a$\sim$d), we are able to obtain $d\propto w^{0.61}$ for $h>h_s$ in Fig. 5(e)
which is in line with the consensus that there  exists a power law and the exponent is greater than 0, but less than 1 for complex meteorite craters \cite{cite24}. However, this consistency is alarming because our data points for $h>h_s$ in fact consist of several constant functions of $w$. In other words, the non-zero exponent in Fig. 5(e) is spurious because it is an artifact of compiling data from different $T$ and $R$, which inevitably is the case for meteorite scholars.


\section{Conclusion and Discussions}
In contrast to solid and liquid projectiles used in previous studies of impact experiments, we adopt melted tin to better simulate the high temperature involved in real meteorite events. Modifying tin temperature enables us to discern V- and U-shape  craters, analogous to the distinction of simple and complex craters in real events. Similar to previous research, we also found the crater width to follow a power-law relation with the impact energy, except that  low and high-energy impacts turn out to exhibit different exponents. The fact that the correlation of depth and  width  is lost as the depth saturates  at high energies allows us to uncover a possible misinterpretation when combining data from different parameters, as is inevitable in field studies. Specifically, we were misled to conclude that depth vs width follows a power law when data from different tin diameter and temperature are plotted on the same graph, which in fact consists of several constant functions of $d(w)$.

To aid the analysis and visualization of the complex interactions between melted tin and granules, we employed a couple of technical methods, including deep learning and image processing.
Mapping the complex patterns generated by tin in its phase transition to a network allows us to  quantitatively compare and elucidate the solidification process. 
We expect our conclusions are not unique to  the use of melted tin, but general to other substances that experience a similar phase transition during the impact process, such as thermoplastic materials.
It is our hope that the advantages of recruiting such a material will encourage more applications. For instance,
the fluid-solidification mechanisms in porous granules, which are highly applicable to additive manufacturing or metal production processes \cite{cite25} and extreme ultraviolet (EUV) lithography in chip production \cite{cite26}, where molten tin is used to generate EUV radiation. In a nutshell, our work is not limited to academic interests, but of potential industrial applications.

\end{document}